# "Crash Test Dummies" for AI-Enabled Clinical Assessment:
## Validating Virtual Patient Scenarios with Virtual Learners

Brian Gin*, University of Illinois Chicago, 0000-0001-7655-3750
Ahreum Lim, University of Illinois Chicago, 0000-0003-3915-2194
Flavia Silva e Oliveira, Federal University of Catalão, 0000-0003-2456-5930
Kuan Xing, University of Iowa, 0000-0001-8309-4804
Xiaomei Song, Case Western Reserve University, 0000-0002-4444-1559
Lukas Shum-Tim, McMaster University, 0009-0001-5261-0332
Gayana Amiyangoda, University of Peradeniya, 0000-0002-7448-818X
Thilanka Seneviratne, University of Peradeniya, Sri Lanka, 0000-0002-4311-3668
Alison F. Doubleday, University of Illinois Chicago, 0000-0003-3941-3016
Ananya Gangopadhyaya, University of Illinois Chicago, 0000-0002-2524-3696
Bob Kiser, University of Illinois Chicago
Dhruva Patel, University of Illinois Chicago, 0000-0001-7973-2754
Kosala Marambe, University of Peradeniya, Sri Lanka, 0000-0002-0847-9103
Lauren Maggio, University of Illinois Chicago, 0000-0002-2997-6133
Ara Tekian, University of Illinois Chicago, 0000-0002-9252-1588
Yoon Soo Park, University of Illinois Chicago, 0000-0001-8583-4335

Correspondence:
Brian Gin, bgin@uic.edu

## Abstract:

**Background:** Artificial intelligence (AI) is increasingly used to assess learners' clinical competencies in virtual standardized patient encounters. However, most approaches evaluate AI assessment using interrater reliability alone and lack a principled measurement framework or an extensible platform for studying case design, learner behavior, and rater decisions jointly shape assessment outcomes. Without such structure, it remains unclear when AI-based assessment is robust, generalizable, and trustworthy. At the same time, collecting data from learners using unvalidated AI learning/assessment systems places them at risk of AI misguidance. We propose a solution to this conundrum by developing AI-based simulated learners that can pre-stress emerging AI learning platforms prior to human learner interaction, facilitating psychometric characterization and key validation steps prior to human exposure.

**Objective:** To develop an open-source open-access, AI-enabled virtual patient platform and accompanying measurement model that together enable robust, generalizable evaluation of clinical competencies across diverse cases and rating conditions.

**Methods:** We built an open-source agent-based virtual patient platform comprising three interacting components: (1) virtual patients designed around clinically authentic scenarios, (2) virtual learners with tunable competency profiles aligned to ACGME core competencies, and (3) multiple independent AI raters that score learner–patient interactions using structured assessment items derived from validated instruments and Key-Features frameworks. Learner-patient encounters generated natural-language transcripts, which were evaluated using a Bayesian measurement model modified from a hierarchical rater-mediated signal detection theory (HRM-SDT) model that treats ratings as decisions under uncertainty, explicitly separating learner ability, case-specific performance states, and rater decision behavior. Model parameters were estimated using Markov chain Monte Carlo methods.

**Results:** Within this integrated platform, the assessment model recovered simulated learners' underlying competency profiles with statistically significant correlations to the competencies used to generate learner behavior across all ACGME domains, despite a highly non-deterministic simulation pipeline. The model enabled estimation of case difficulty by competency, revealing clinically interpretable differences across virtual patient scenarios. Rater sensitivity (*detection*) and severity/leniency (*criteria* thresholds) were stable across AI raters using identical models and prompts but different random seeds, supporting the internal consistency and robustness of the platform. We used our findings to propose a generalizable "safety blueprint" for deploying AI tools with learners, based on staged achievement of entrustment-based validation milestones.

**Conclusions:** Developing AI assessment within a sound psychometric measurement framework and a purpose-built virtual patient platform enables robust, interpretable, and generalizable estimation of clinical competencies. By using AI learners as "crash test dummies" to stress-test assessment systems, this work provides an open, extensible foundation for studying and validating AI-assisted clinical assessment prior to deployment with human learners and raters.

**Introduction:**

Health professions educators are actively exploring the ability of artificial intelligence (AI) large language models (LLMs) to model human dialog interactively as a potential way to lower the resources needed to both design and implement virtual standardized patient cases.[1–3] When combined with automated learner feedback, AI-driven virtual patient scenarios present an opportunity to increase access to both formative and summative assessment – serving in roles ranging from interactive textbook to full-fledged OSCE station replacement. While there are many aspects of human-human interaction that educators may not expect (nor desire) AI mimicry to replace, AI-driven learner chatbots and AI-driven clinical supervisors may provide learners with supplemental, self-paced opportunities for deliberate practice in low-stakes environments. Further, the ability of AI to provide individualized feedback on learners' individual case performance, as well as their longitudinal learning trajectories has been envisioned by educators as filling an unmet need.[4,5]

From an assessment standpoint, all these aspirations hinge critically on implicit (and often unproven) assumptions about validity – that the AI acting in the role of a patient is consistent and standardized, that the AI acting in the role of a rater or supervisor is accurate and precise, and that together they are fair and free from bias towards the learner. Yet, most empirical work to date has been unable to answer the central measurement question: **when do AI-based virtual standardized patient encounters yield robust, generalizable estimates of clinical competence?** Without quantitative performance metrics for the cases themselves – and for the scoring systems coupled to those cases – we cannot determine how consistently a case functions, which competencies it meaningfully elicits, or whether automated ratings and feedback reflect stable underlying competence rather than idiosyncrasies of the interaction.

Existing evaluations of AI-driven virtual patient assessments generally fall into three categories. First, studies assess the interaction experience (e.g., satisfaction, perceived realism, educational utility).[6–10] These perceptions are informative but primarily reflect learners' explanations of how they engaged with the system – closely aligned with *response-process* evidence.[11] Second, studies assess the feedback or ratings produced by AI by comparing them to human judgments, often emphasizing interrater reliability or overlap in feedback themes[12–15] – relating to *both internal structure* and *relation to other variables*. However, such comparisons provide limited evidence for validity, because agreement with humans is not itself a psychometric standard – human ratings can be inconsistent, and reliability alone does not establish a defensible measurement structure or generalizability across cases, items, and raters. Third, a smaller body of work examines learning outcomes,[16–18] including early randomized trials;[19,20] these are important (i.e. for *consequential validity*) but do not resolve the upstream question of whether the assessment signal produced by the case-and-rater system is stable and interpretable. Across these approaches, a common limitation remains: we lack an extensible framework that can quantify how case design, learner behavior, and rater decisions jointly shape observed scores and inferred competence.

However, limitations to data collection create an additional barrier to such validation – and for good reason: involving learners as study subjects puts them at risk to AI inconsistencies, inaccuracies, and biases.[21,22] Even if the learners are aware that an AI-based assessment is experimental, their learning can still be influenced by the experience, especially for learners who are in early, formative stages. A subtly misrepresented disease process, or a social concern that is represented more as a stereotype than with authentic nuance can frame learning unfairly. Feedback from an AI supervisor does not need to be inaccurate to be misleading, as indirect omissions and over-emphases can also skew learner perceptions. As such, it becomes difficult (unacceptably high risk) to collect the data necessary to perform a rigorous quantitative validation of AI-based virtual patient cases.

To address this gap, we propose an intermediate solution: **evaluate virtual patient cases and automated assessment systems before involving human learners, by using AI-simulated learners as "crash test dummies"** – i.e., AI agents designed to mimic the clinical communications and reasoning behaviors of human learners. In this approach, a pool of virtual learners is constructed to represent a wide range of realistic competency profiles. The same learners can be deployed repeatedly across multiple virtual patient cases and evaluated by multiple independent AI raters. This design makes it possible to generate the structured, repeated-measures data needed to estimate psychometric properties of (a) the learner–case encounters, (b) the assessment items attached to those encounters, and (c) rater decision behavior – while holding constant the underlying competency profile used to generate learner behavior. In turn, this enables performance metrics that are typically unavailable in early-stage AI assessment research: how difficult each case is by competency, which competencies each case elicits well (or poorly), and how rater sensitivity (*detection*) and severity/leniency (*criteria*) vary across competencies and rating conditions.

Critically, this framework also provides a way to separate sources of variability that are otherwise conflated in AI assessment: variability from the learner's competence, from the clinical context and case design, and from the rater's decision process under uncertainty. By treating ratings as decisions rather than direct measurements, we can move beyond agreement-with-humans as the dominant evaluation target and instead ask whether an AI assessment system yields stable, interpretable, and generalizable competence estimates across cases and raters – conditions that are necessary for trustworthy formative feedback and, eventually, defensible summative assessment.

**AIMS:**

AIM 1. Develop and evaluate a measurement framework that enables robust estimation of learners' clinical competencies on a common scale across virtual patient cases and AI raters.
AIM 2. Estimate case difficulty by competency, quantifying how different virtual patient scenarios differentially elicit (and challenge) specific ACGME-aligned competencies.[23]
AIM 3. Estimate competency-specific rater tendencies (e.g., sensitivity and severity/leniency), enabling adjustment for model-, prompt-, and rater-dependent rating behavior as AI systems evolve.

To operationalize these aims, we built an open-source, agent-based platform that can (1) enact validated virtual patient cases, (2) generate repeatable "virtual learner" performances spanning ACGME-aligned competency profiles, (3) apply multiple independent AI raters to each transcript, and (4) estimate learner-, case-, and rater-level parameters using a novel measurement model.

## Methods:

**Overview and study design.** We conducted a simulation-based psychometric evaluation in which a fixed pool of AI-simulated learners interacted with multiple AI virtual patient cases. Each resulting transcript was independently scored by multiple AI raters, and the full set of ordinal ratings was fit with a two-layer measurement model to recover: (a) learner competence by ACGME domain, (b) case difficulty by competency, and (c) rater tendencies (detection/criteria), including competency-specific rater effects (Figure M1).

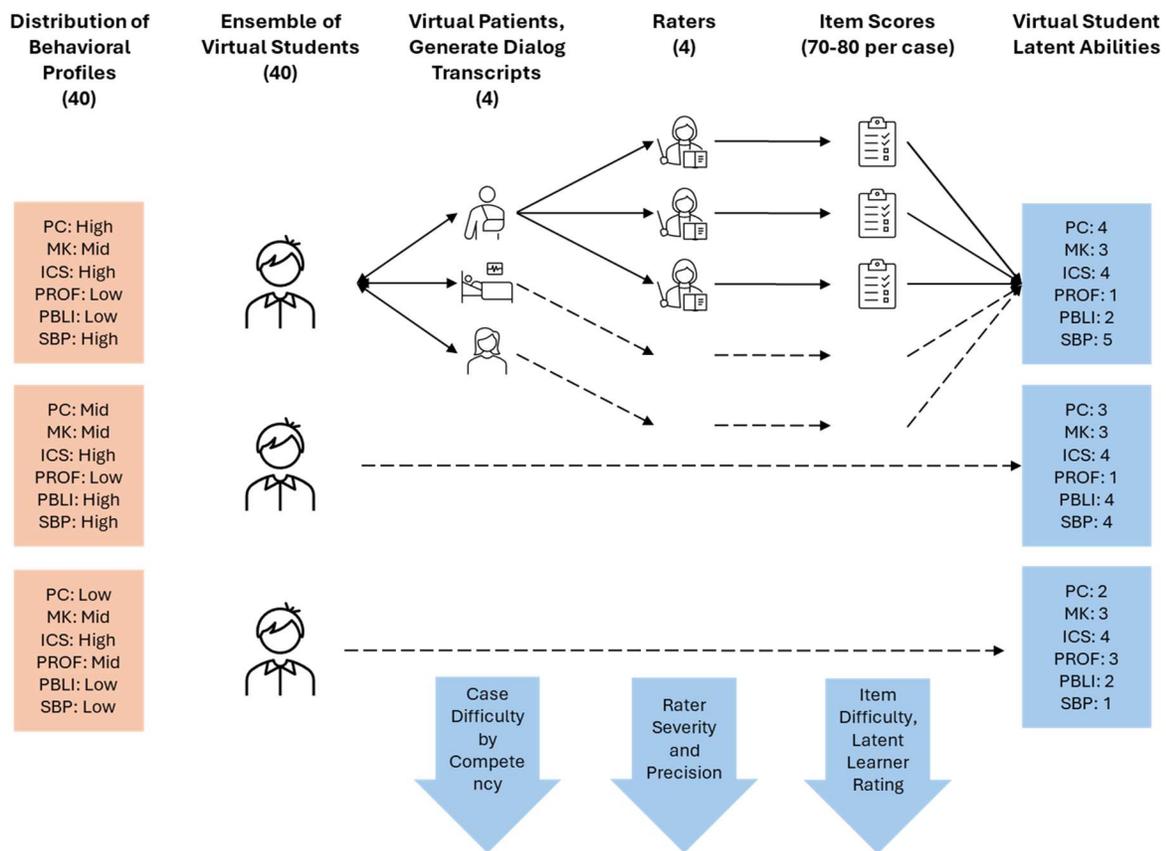

*Figure M1: Overview of study design, using virtual students to interact with virtual patients, scored by virtual raters. The overall goal is to collect validity evidence for a measurement model that assesses each student's ACGME competencies based on their virtual patient interactions. This also requires assessing properties of the virtual patient cases themselves (case difficulty by competency), the AI-raters (detection and criteria), and the items used to assess the learners (difficulty and item characteristic curves). Abbreviations: PC=patient care; MK=medical knowledge, ICS=interpersonal and communication skills; PROF=professionalism; PBLI=practice based-learning and improvement; SBP=systems-based practice.*

**AI-enabled virtual patient platform.** We developed an agentic virtual patient system in which multiple LLM/LRM agents jointly enact a virtual patient encounter that can be run with either a human learner (voice or text) or a virtual learner (text). The platform automatically logs the full learner–patient interaction as a transcript for downstream rating and analysis.

**Virtual patient cases.** We instantiated four virtual patient cases (one virtual patient per case). Cases were authored by humans and previously validated in their original educational contexts. The four scenarios were designed to elicit distinct clinical and communication demands while sharing a common competence framework. In each case, learners were expected to (i) elicit key history elements, (ii) demonstrate appropriate clinical reasoning and planning concordant with their simulated training level, and (iii) communicate in a patient-centered manner. Each case included instructions to conduct the interaction in two phases: a first history-taking phase, followed by a second phase in which the learner is asked to explain their diagnostic thinking, describe exam components and tests they would perform, synthesize a plan, and provide follow up instructions (Appendix B2.1)

The cases included chief complaints/diagnoses of: Headache/meningitis, Fever of unknown origin/lupus, Colon cancer/goals of care, and Back pain/multimodal treatment (Box 1). Each case was designed to elicit all of six of the ACGME competencies: Patient Care (PC), Medical Knowledge (MK), Interpersonal and Communication Skills (ICS), Professionalism (PROF), Practice-Based Learning and Improvement (PBLI), and Systems-Based Practice (SBP).[24] Each competency was measured by a minimum of 5 primary items (see below).

---

***Box 1:*** *Case scenario summaries and learner tasks (`abbreviated name` in parentheses)*

Back Pain (`BackPain`): A 55-year-old elementary school teacher presented for follow-up with 6 months of worsening low back pain and persistent fatigue. Learners were expected to screen for safety-critical red flags, differentiate mechanical versus inflammatory features, propose an initial evaluation and conservative management plan, and align next steps with patient-centered goals.

Colon Cancer (`ColonCA`): A 68-year-old retired engineer presented after evaluation revealed metastatic colon adenocarcinoma and sought understanding of prognosis and treatment options. Learners were expected to communicate serious news clearly and empathetically, explain staging and the rationale for further testing, elicit values and preferences, and coordinate next steps (e.g., oncology referral and logistics).

Fever of unknown origin (`Fever`): An 18-year-old college student presented with a month of intermittent fevers and fatigue that progressed to daily fevers with dyspnea, chest pain, edema, and arthralgias. Learners were expected to assess acuity, generate a prioritized differential (infectious, autoimmune, malignancy), integrate findings suggestive of systemic lupus erythematosus and nephritis, and outline an appropriate diagnostic and escalation plan.

Meningitis (`Meningitis`): A 30-year-old graduate student presented to the emergency department with high fever, rapidly worsening headache, photophobia, neck stiffness, and a petechial rash concerning for meningococcal meningitis. Learners were expected to recognize a time-critical emergency, initiate appropriate precautions and urgent evaluation, begin empiric

---

treatment, reconcile antibiotic allergies, and consider systems-level actions (public health notification and contact prophylaxis planning).

**Case extraction and retrieval-augmented case grounding.** To make cases executable by the virtual patient agents, we used an automated extraction step that converts each human-authored case into (a) structured prompting materials and (b) a case-specific retrieval-augmented generation (RAG) database[25] used to maintain grounding and consistency during the encounter (Appendix B.1).

**Virtual learners and competency profiles.** We developed a virtual learner agent whose behaviors can be tuned to represent different realistic learner skill sets (Figure M2). For this study, we modeled the learner as a first-year medical student and parameterized competence using the ACGME core competencies. Competency behavioral anchors (with definitions adapted to medical students) were used to steer the learner's behavior during the encounter (Appendix B.2).[26,27] Learner competency values were sampled from a random distribution and compiled into custom prompts; a secondary LLM-based control layer adjusted outputs at inference time when the generated behaviors did not sufficiently reflect the intended anchors. In the current implementation, we generated a pool of 40 virtual learners deployed across cases.

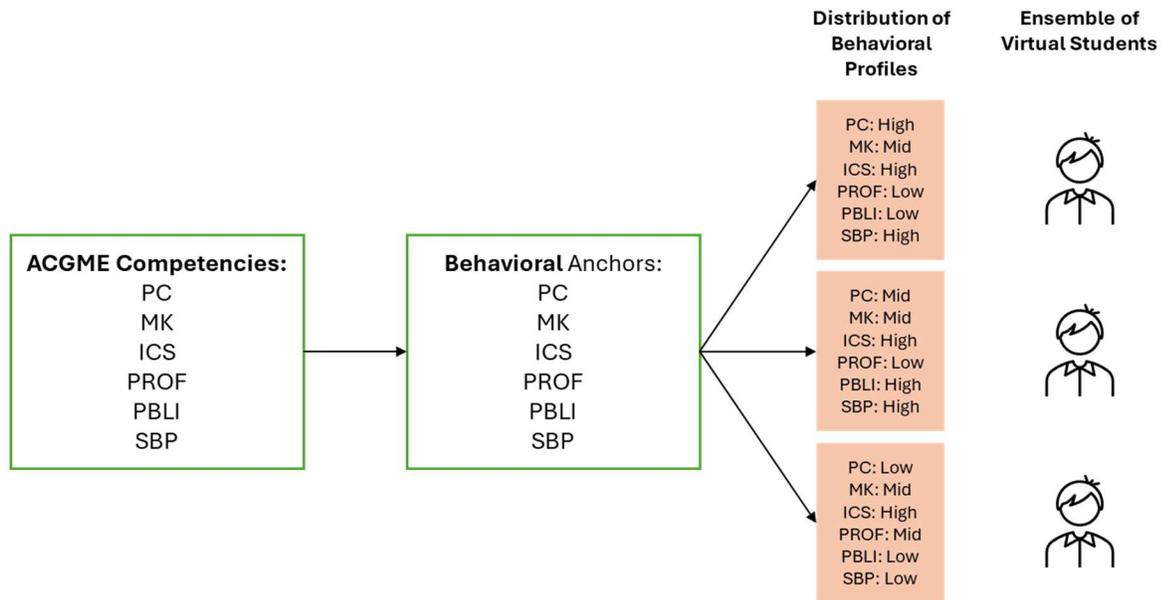

*Figure M2: Generating an virtual "cohort" of AI-enacted virtual students to exhibit different ACGME competency profiles based on the behavioral anchor descriptions for each competency. Abbreviations: PC=patient care; MK=medical knowledge, ICS=interpersonal and communication skills; PROF=professionalism; PBLI=practice based-learning and improvement; SBP=systems-based practice.*

**Assessment item generation and competency mapping.** Each case's scoring instrument (Appendix B.4) was assembled from multiple sources: (1) previously-validated items including a global communications rating scale and existing case rubrics, and (2) *de novo* AI-generated items to improve competency coverage. *De novo* item generation was guided by an LLM-based Key

Features extraction step inspired by Bordage's Key Features framework,[28] followed by item-writing from those extracted key features and an additional pipeline that targeted gaps in ACGME competency coverage using case data plus ACGME definitions.[27] The measurement model estimated a six-dimensional competency profile (Theta1-Theta6), which we mapped to ACGME-aligned competency domains as follows: Theta1 = PC, Theta2 = MK, Theta3 = ICS, Theta4 = PROF, Theta5 = PBLI, and Theta6 = SBP. Finally, we developed a weighting map linking each item to one or more ACGME competencies (item–competency "theta loadings"), which were then treated as fixed design inputs to the measurement model.

**Virtual raters and rating design.** We implemented a virtual rater agent that scores each transcript using the item descriptions and any additional case context (Appendix B.3). To support estimation of rater tendencies, we used multiple AI raters per transcript (N = 4), allowing rater detection and criteria (and optional item-specificity) to vary.

**Learner-patient transcript generation workflow.** For each simulated learner, we generated one transcript per case by running the learner agent against each virtual patient agent. Each transcript was then independently scored by four AI raters on the full item set for that case, yielding repeated ordinal ratings suitable for hierarchical estimation of learner-, rater-, item-, and case-level effects.

**Measurement model:**

*Modified HRM-SDT model*
We specify a Bayesian hierarchical rater-mediated signal detection theory (HRM–SDT) model for ordinal ratings (Figure 1ab).[28] Unlike common rater-effect IRT approaches that primarily summarize rater differences as severity (and treat scale-use implicitly), HRM-SDT considers the process of rater decision-making, modeling both rater *detection* and *criteria* separately.[29] By contrast, G-theory would quantify rater-related variance and support sampling decisions, but it does not model the evidence-to-category mechanism that distinguishes threshold shifts from reduced detection. Even with AI raters, repeated ratings can differ systematically due to prompt sensitivity or rubric interpretation; modeling these differences improves identifiability and prevents rater idiosyncrasies from being mistaken for learner or case effects. Thus, our model separates (i) examinee competency, (ii) item/case difficulty and thresholds, and (iii) rater detection and criteria. The likelihood is defined by a hierarchical joint probability model, which is used for inference from observed ratings (Appendix A). Modifications include the additions of: case difficulty parameters, competency-dependence (of case difficulty and rater parameters), and an applicability (opportunity) gate to prevent unscored items (due to virtual patient variations) from biasing learner competency estimates.

*Indices*
- $i$ examinee/learner ($i = 1, \ldots, N$)
- $j$ rater ($j = 1, \ldots, J$)
- $l$ item ($l = 1, \ldots, L$)
- $p$ competency dimension ($p = 1, \ldots, P$)
- $k$ category boundary ($k = 1, \ldots, 4$)

*Observed ratings*
For each $(i, j, l)$, the observed rating is ordinal:
$$Y_{ijl} \in \{1,2,3,4,5\}.$$

*Latent performance states*
For each $(i, l)$, let $\eta_{il} \in \{1,2,3,4,5\}$ denote a discrete latent performance category (the "true" level on the rubric for item $l$ for examinee $i$). We do not sample $\eta_{il}$ directly; instead, it is marginalized in the likelihood.

*Competency dimensions and fixed item loadings*
Each examinee has a $P$-dimensional competency vector $\boldsymbol{\theta}_i \in \mathbb{R}^P$. Each item $l$ has a *known* loading vector $\mathbf{a}_l \in \mathbb{R}^P$ describing how competency dimensions contribute to item performance (e.g., from a competency blueprint or rubric design). For identifiability and interpretability, assume $\mathbf{a}_l$ is normalized to unit length.

*Case membership and theta-groups*
Each item belongs to a case $q(l) \in \{1, \dots, C\}$. Each item is also assigned to a theta-group $g(l) \in \{1, \dots, G\}$ (e.g., the primary competency dimension for that item), used to allow rater behavior to vary across item types.

*Observed applicability (opportunity) indicators.*
Not all items are meaningfully scorable in every transcript/case context: some skills may not be elicited by the virtual patient during the conversation, and "missing" ratings should not be conflated with poor performance. We therefore introduce a Bernoulli applicability indicator (Figure 1b),
$$A_{ijl} \in \{0,1\},$$
generated from a *case×theta-group* logit $\omega_{q(l),g(l)}$ via a logistic link,
$$\mathrm{Pr}\big(A_{ijl} = 1\big) = logit^{-1}(\omega_{q(l),g(l)}),$$
$A_{ijl} = 1$ indicates that item $l$ was applicable (the examinee had an opportunity to demonstrate the behavior/skill) and thus can contribute a rating, and $A_{ijl} = 0$ indicates structural non-applicability. When $A_{ijl} = 0$, the corresponding rating $Y_{ijl}$ is treated as unobserved (i.e., excluded from the rating likelihood); when $A_{ijl} = 1$, the rating follows the HRM–SDT likelihood.

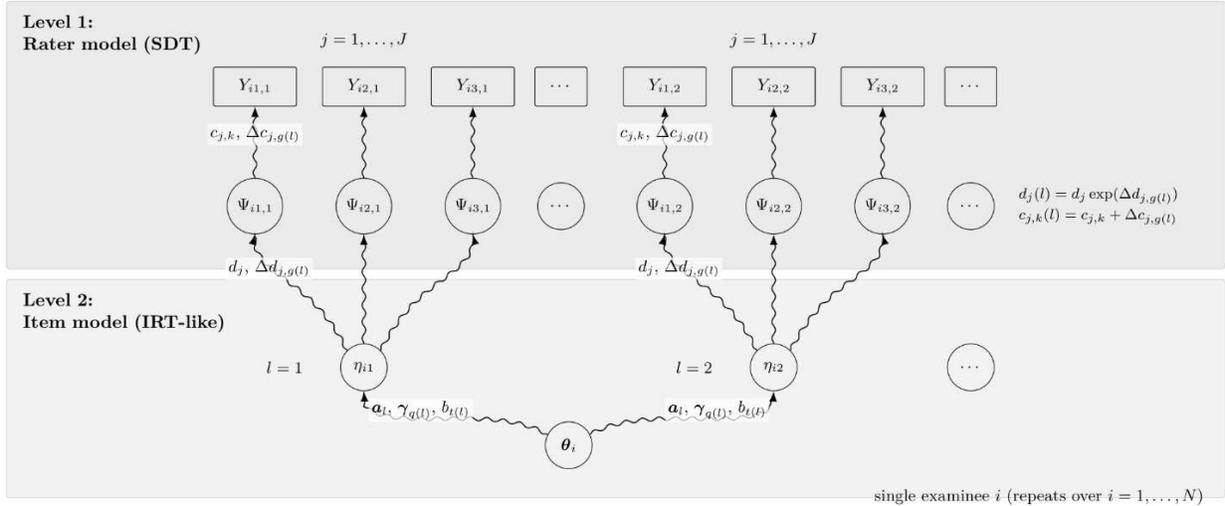

*Figure 1a: HRM–SDT model for transcript-based ordinal ratings. Learner competency and case difficulty generate a latent "true" rubric level per item $\eta_{il}$ (Stage 1). Each rater then produces an internal evidence variable $\Psi_{ijl}$ and assigns an observed score $Y_{ijl}$ by applying rater-specific decision thresholds (criteria, $c_{j,k}$) and signal detection (sensitivity/precision/discrimination, $d_j$) (Stage 2). Modeling both components distinguishes raters who are systematically stringent/lenient (shifted criteria thresholds) from raters who are less able to separate performance levels from transcript cues (low sensitivity). The wavy arrows indicate nonlinear relationships between variables.*

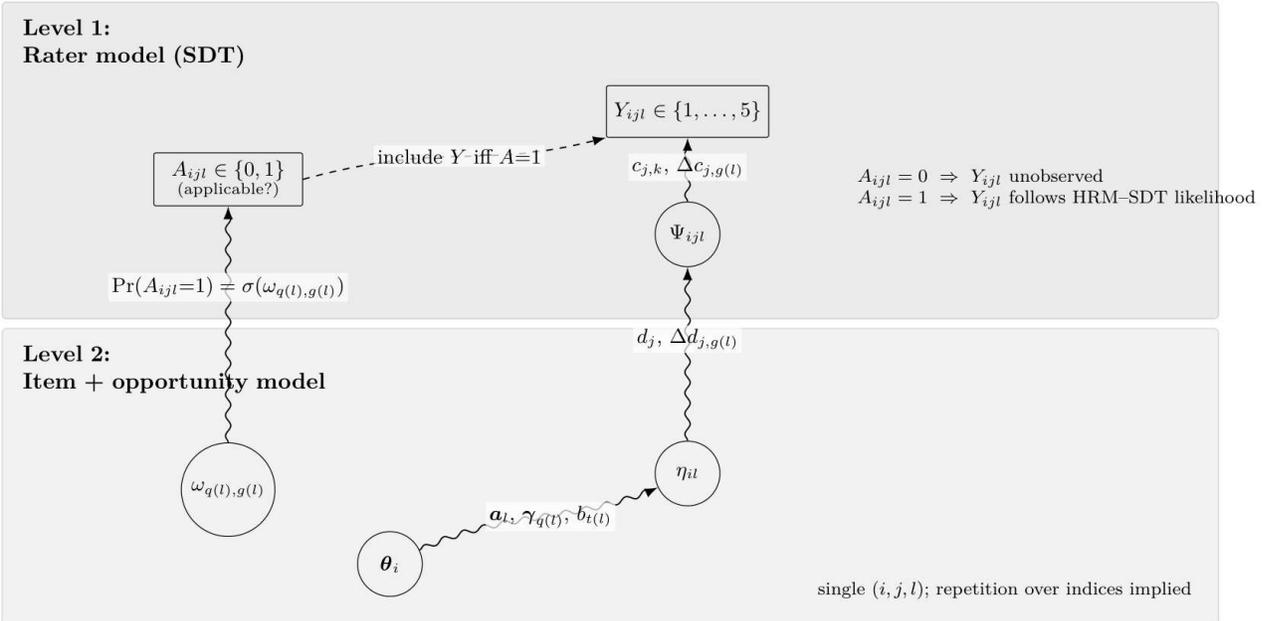

*Figure 1b. Applicability (opportunity) gate illustrated for a single representative rating. When an item is structurally non-applicable in a transcript (no opportunity to observe the skill), the rating is treated as unobserved so that missingness is not conflated with poor performance; when applicable, the rating follows the HRM–SDT likelihood.*

*Rater parameters and their competency-group dependence*

HRM–SDT distinguishes two key aspects of rater behavior: *detection* (often referred to as *discrimination*, *sensitivity*, or *precision*) and *criteria* (or *thresholds*). *Detection* reflects how well a rater can differentiate between performance levels; higher detection implies steeper separation of response probabilities across levels. *Criteria* are the rater's category thresholds on the latent evidence scale that map internal evidence to the 5-point rubric. Shifting criteria upward requires more evidence to award higher categories and is often described as increased *severity* (and downward shifts as *leniency*); however, criteria more generally capture rater scale use (i.e., where each category boundary is placed), not only overall stringency.

Each rater $j$ is characterized by a *detection* (sensitivity) parameter $d_j > 0$ and a set of ordered *criteria* (decision thresholds) $c_{j,1} < \cdots < c_{j,4}$. In principle, HRM–SDT can allow fully item-specific rater parameters (e.g., $d_{j,l}$ and $c_{j,k,l}$) to capture rater-by-item differences in rubric interpretation; however, to balance identifiability and interpretability given the limited number of raters and repeated ratings per transcript, we allow rater behavior to vary at the competency-group level $g(l)$ rather than be fully item-specific. To allow rater behavior to vary across item types, both sensitivity and criteria may depend on an item's competency-group $g(l)$:

$$d_j(l) = d_j \exp\big(\Delta d_{j,g(l)}\big), \qquad c_{j,k}(l) = c_{j,k} + \Delta c_{j,g(l)}.$$

Detection is modeled multiplicatively to enforce positivity and to interpret competency-group effects as proportional changes in sensitivity. Criteria are modeled additively because they live on the latent evidence scale of $\Psi_{ijl}$, where shifts correspond directly to severity/leniency offsets (and preserve ordering given ordered baseline criteria).

*Relative case difficulty*

We allow cases to shift item difficulty in a competency-specific (i.e. dimension-specific) manner via $\boldsymbol{\gamma}_q \in \mathbb{R}^P$. To ensure identifiability, we mean-center case effects across cases *within each dimension*. For case difficulty shifts to be interpretable as meaningful differences between cases, there must be shared items (with associated shared scoring thresholds) across cases. Therefore, we included a subset of universal items that are reused across multiple cases and scored with the same rubric. We jointly estimated these items' shared scoring thresholds across all cases.

*Two-stage structure (IRT-like → SDT)*

The model is most naturally described in two linked stages:

1. **Stage 1 (IRT-like item model).** Learner competency and case effects determine the distribution of the latent performance state $\eta_{il}$. Specifically, an ordered-logit model maps an item-specific linear predictor $s_{il} = \boldsymbol{\theta}_i^\top \mathbf{a}_l + \boldsymbol{\gamma}_{q(l)}^\top \mathbf{a}_l$ to cumulative probabilities of $\eta_{il}$ using ordered thresholds $\{b_{t(l),k}\}_{k=1}^{4}$. This plays the role of an ordinal IRT model with known item loadings $\mathbf{a}_l$ and estimated step/transition parameters $b_{t(l),k}$.

2. **Stage 2 (SDT rater model).** Conditional on $\eta_{il}$, each rater $j$ generates an internal evidence/decision variable $\Psi_{ijl}$ whose location is proportional to (a normalized version of) $\eta_{il}$ and whose noise is logistic. The observed ordinal rating $Y_{ijl}$ is produced by discretizing $\Psi_{ijl}$ with the rater's criteria $c_{j,k}(l)$, yielding an ordered-logit form for $\Pr(Y_{ijl} \leq k \mid \eta_{il})$ in which sensitivity enters through $d_j(l)$ and severity/leniency enters through $c_{j,k}(l)$.

For inference, the discrete latent states $\eta_{il}$ are marginalized, so each observed rating contributes a mixture likelihood over the five possible latent performance levels.

*Likelihood (HRM–SDT)*
Let $\sigma(x) = (1 + e^{-x})^{-1}$ be the logistic CDF.

*Stage 1: competency → latent performance state*
Define the Stage 1 linear predictor

$$s_{i,l} = \boldsymbol{\theta}_i^\top \mathbf{a}_l + \boldsymbol{\gamma}_{q(l)}^\top \mathbf{a}_l.$$

The latent performance state $\eta_{il}$ follows an ordered-logit model:

$$\Pr(\eta_{i,l} \leq k \mid \boldsymbol{\theta}, \boldsymbol{\gamma}, b) = \sigma(b_{t(l),k} - s_{i,l}), \quad k = 1, \dots, 4.$$

*Stage 2: latent performance → rater score (SDT layer)*
Define a normalized latent performance level $\tilde{\eta}_{il} \in [0,1]$ by

$$\tilde{\eta}_{il} = \frac{\eta_{il} - 1}{4}.$$

Let the rater decision variable be

$$\Psi_{i,j,l} = d_j(l) \cdot \tilde{\eta}_{i,l} + \epsilon, \quad \epsilon \sim \text{Logistic}(0,1).$$

The observed score is produced by discretizing $\Psi_{ijl}$ using the rater's criteria. Equivalently, integrating out $\Psi_{ijl}$ yields an ordered-logit form:

$$\Pr(Y_{ijl} \leq k \mid \eta_{i,l}) = \sigma(c_{j,k}(l) - d_j(l) \cdot \tilde{\eta}_{i,l}), \quad k = 1, \dots, 4.$$

*Inference/estimation*
We performed posterior inference using Hamiltonian Monte Carlo[30] with the No-U-Turn Sampler (NUTS) implemented in a JAX-backed sampler.[a] Sampling used 4 independent chains, 1000 warmup (tuning) iterations per chain, and 750 retained posterior draws per chain, with a target acceptance probability of 0.99. Convergence was assessed using standard diagnostics (e.g., $\hat{R}$, effective sample size, and visual inspection of trace plots). The discrete latent states $\eta_{il}$ were marginalized, yielding a differentiable likelihood suitable for gradient-based HMC. Hyperparameters and priors are detailed in Appendix A.

## Results:

**Overview.** We summarized post-estimation results from the fitted Bayesian HRM-SDT measurement model in a sequence that mirrored the model's inferential goals: (i) recovery of simulated learner competencies, (ii) empirical separability of competency dimensions, (iii) theta-specific case effects (relative case difficulty/easiness), (iv) cross-case stability diagnostics (between-learners and within-learners), and (v) rater behavior (theta-dependent detection and criteria). Because the study used simulated AI learners with known generating competencies, we interpreted recovery and correlation diagnostics relative to the ground-truth design.

---

[a] Python 3.12.9, PyMC 5.27.0, JAX 0.8.2

**Learner competency recovery.** Estimated learner competencies were positively correlated (Figure 2) with the generating (truth) competencies across all six competency dimensions (Pearson r ranged from 0.38 to 0.76; all p<0.05). Recovery was strongest for Theta1 (PC) (r=0.76) and Theta4 (PROF), and it was weakest for Theta5 (PBLI) (r=0.38). This pattern suggested that some competencies were more consistently elicited and scored across cases and raters than others within the current simulation design.

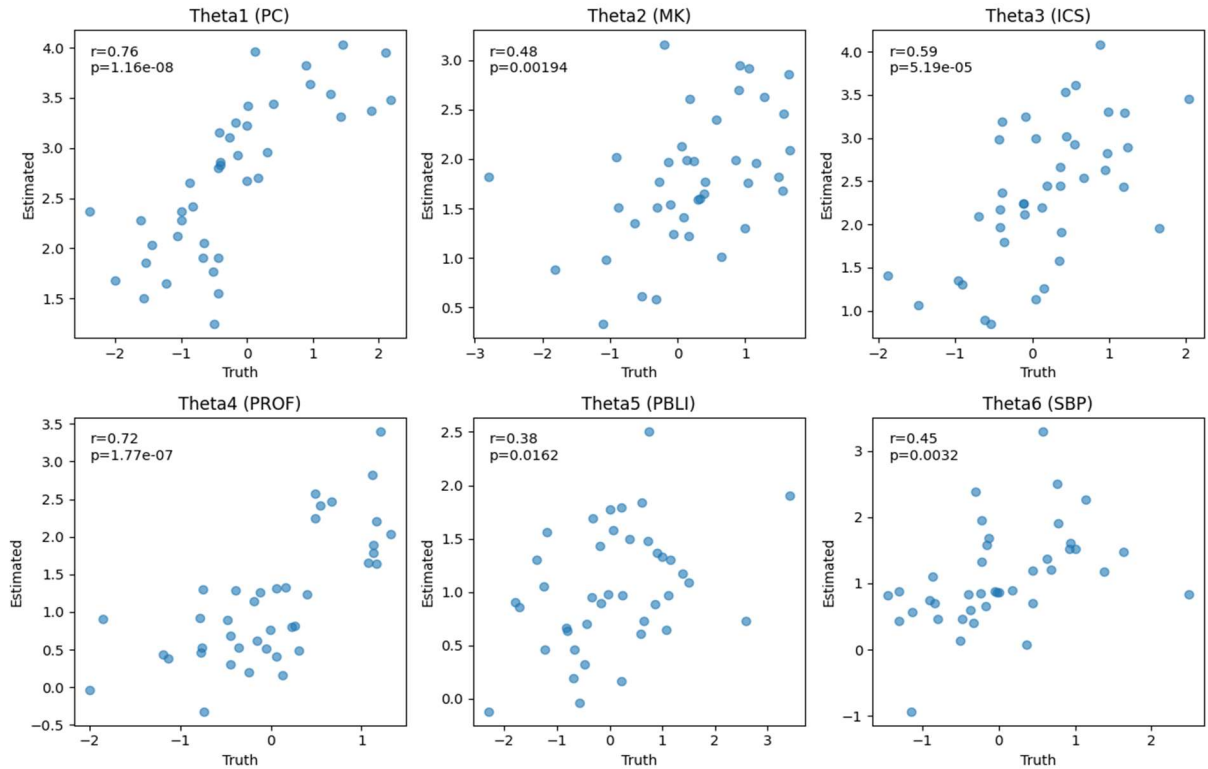

*Figure 2. Correlation between true and estimated competencies by dimension (simulation).*

The heatmap comparison (Figure 3) further indicated that many learners' multivariate competency profiles were recovered in the correct relative ordering, while still exhibiting dimension-specific shrinkage and residual error.

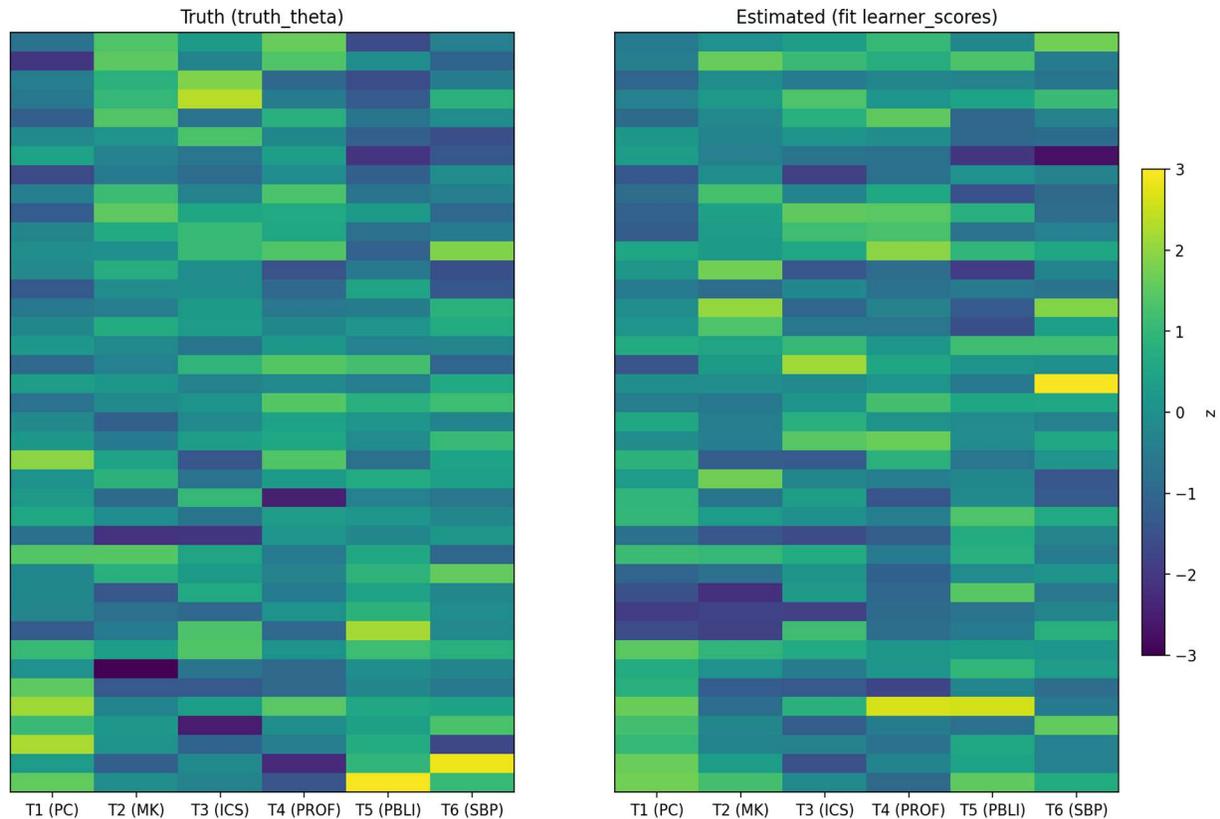

*Figure 3. Heatmap comparison of true vs estimated competency profiles across learners (simulation).*

**Separability of competency dimensions (across learners).** We evaluated whether the six ACGME competency dimensions behaved as empirically distinct constructs by examining correlations between theta dimensions across learners. In the simulation design, learner competency coordinates were generated independently (no deliberate cross-dimension correlations within a learner), so large off-diagonal correlations would not have been expected aside from finite-sample variability. Consistent with this, the generating competency correlation matrix (Figure 4) showed modest off-diagonal correlations (max |r|=0.34).

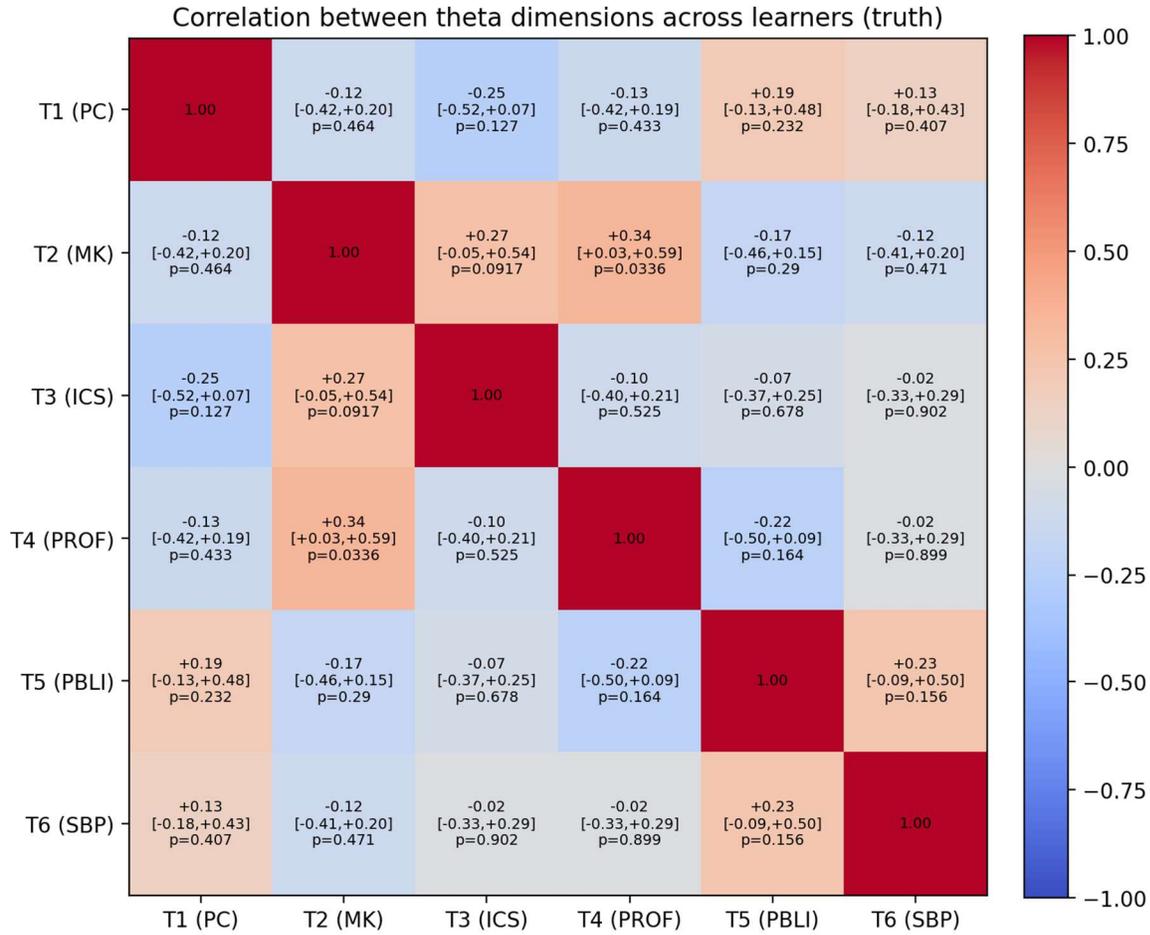

*Figure 4. Correlation between generating competency dimensions across learners (simulation design control).*

The estimated correlation matrix (Figure 5) exhibited somewhat larger off-diagonal structure (max |r|=0.54; strongest estimated association was between Theta3 (ICS) and Theta4 (PROF) with r=0.54). Because the generating dimensions were independent, these induced correlations were interpreted as evidence that some competencies were not fully separable under the current measurement design (e.g., overlapping item information, imperfectly orthogonal loading structure, or posterior coupling due to limited information in some dimensions), analogous to projection/rotation effects in a factor model.

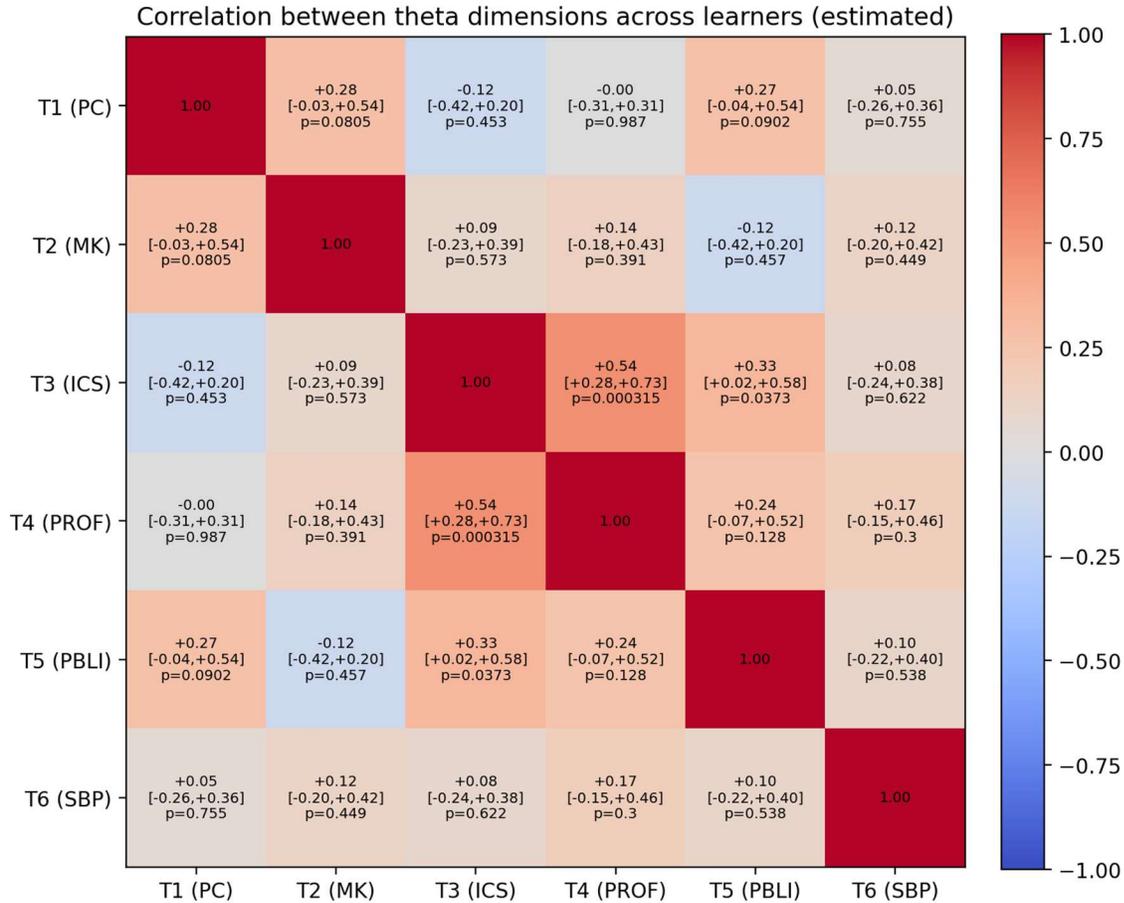

*Figure 5. Correlation between estimated competency dimensions across learners (post-estimation diagnostic).*

**Theta-specific case difficulty/easiness.** We summarized case effects via theta-specific case shifts in the Stage 1 (IRT-like) layer (Figures 6 and 7). After centering per dimension, higher values indicated that a case was relatively easier on items aligned with a given theta, whereas lower values indicated relative difficulty. Across dimensions (averaged naively for description), `Meningitis` and `Fever` tended to be harder (more negative mean shifts), whereas `ColonCA` tended to be easier (more positive mean shifts). Case differences were most pronounced for Theta6 (SBP) (spread across cases was 5.10), with `ColonCA` the easiest (+2.88) and `Meningitis` the hardest (-2.22) on that dimension.

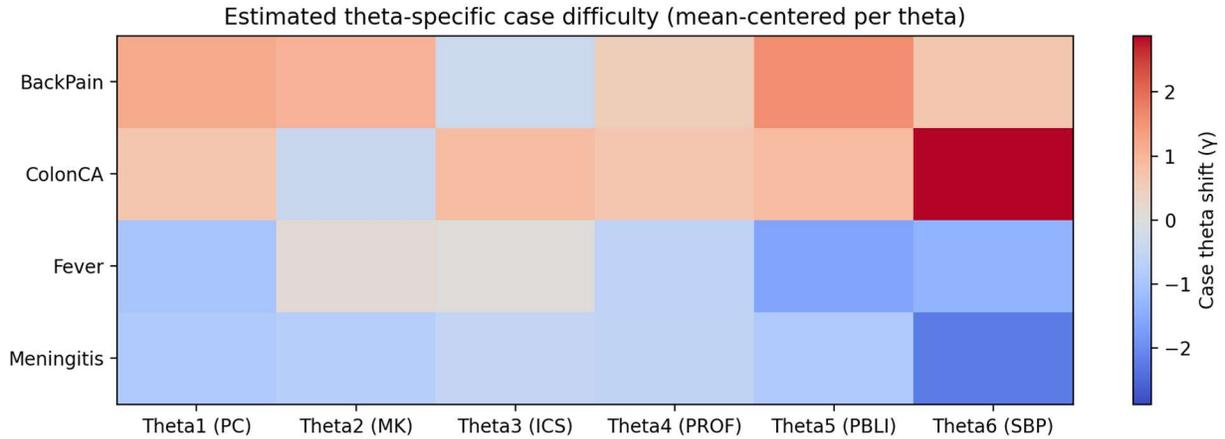

*Figure 6. Estimated theta-specific case shifts (mean-centered per theta).*

Uncertainty intervals (Figure 7) suggested that some case-by-theta contrasts were estimated more precisely than others, consistent with unequal item coverage and loading mass across dimensions.

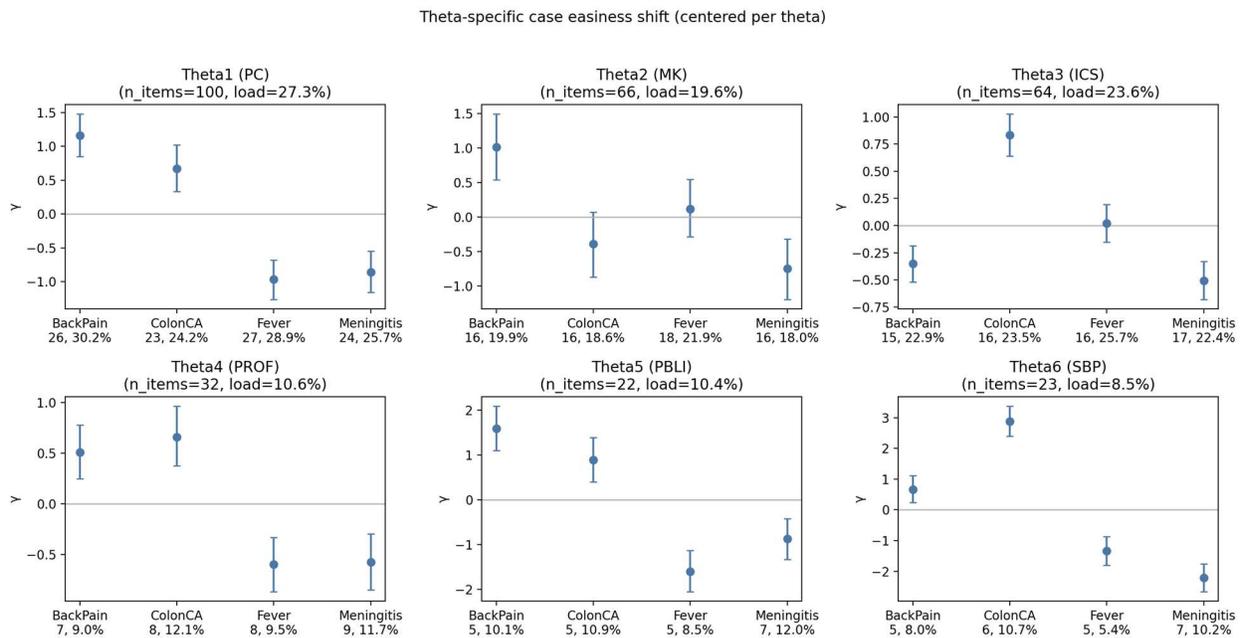

*Figure 7. Theta-specific case shifts with credible intervals and item-coverage summaries.*

**Cross-case consistency (between learners).** To evaluate case invariance of competency estimates, we computed correlations across learners between case-conditional competency estimates for each pair of cases (Figure 8). Average cross-case correlations were strongest for Theta4 (PROF) (mean r=0.65) and also high for Theta1 (PC) and Theta3 (ICS), indicating stable ordering of learners on these competencies across cases. In contrast, Theta2 (MK) (mean r=0.03) and Theta5 (SBP) were near zero and not statistically distinguishable from chance under permutation, suggesting weak identification or inconsistent measurement of those dimensions across cases.

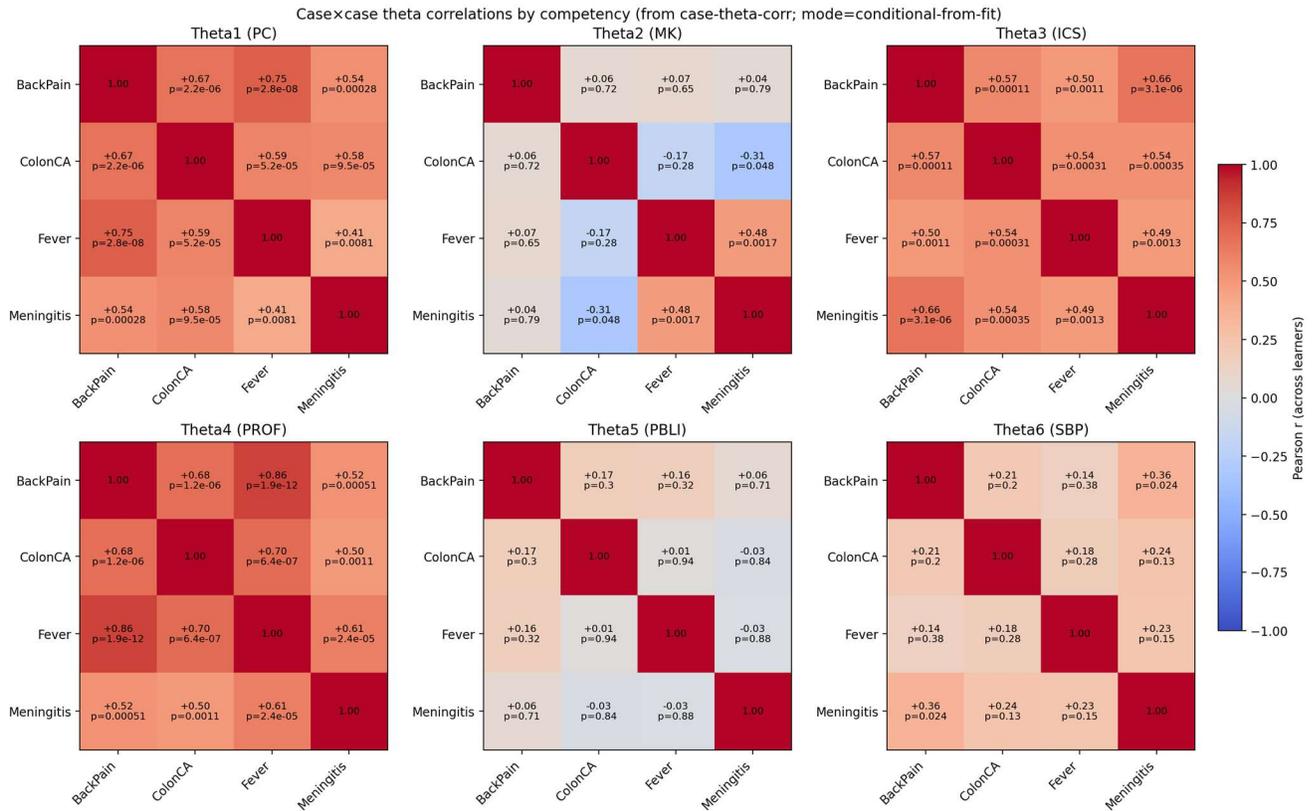

*Figure 8. Between-learner cross-case competency consistency by dimension (Pearson r across learners).*

Aggregating across dimensions yielded moderate cross-case consistency overall (Figure 9). The strongest aggregated case pair was `BackPain` vs `Fever` (r=0.48), whereas the weakest was `ColonCA` vs `Meningitis` (r=0.28). This pattern suggested that some case pairs elicited more comparable evidence about learner competency profiles than others.

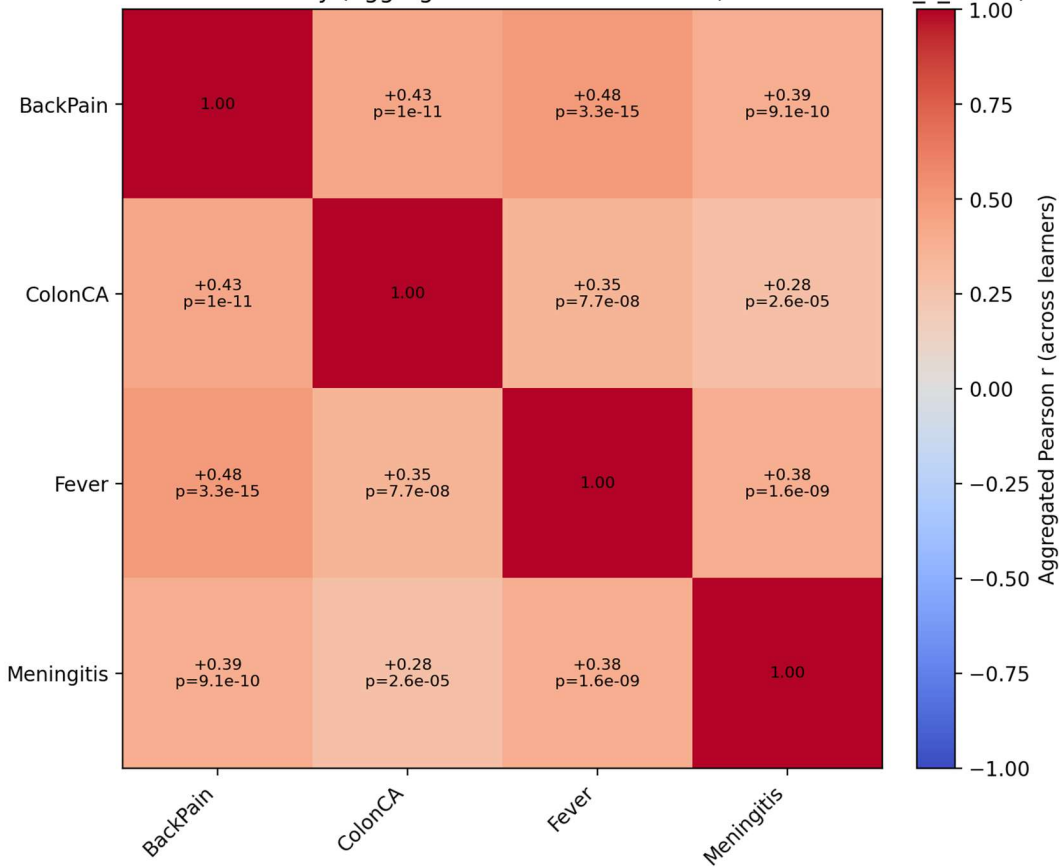

*Figure 9. Aggregated between-learner cross-case competency consistency across dimensions (Fisher-z mean).*

**Cross-case profile consistency (within learners).** Complementing between-learner correlations, we evaluated within-learner stability of multivariate competency profiles across cases by correlating each learner's P-dimensional case-conditional competency vectors across dimensions (Figure 10). Overall, within-learner profile correlations were moderate (mean r=0.57, 95% CI [0.53, 0.61]). The most similar case pair was `Fever` vs `Meningitis` (mean r=0.63), whereas the least similar was `ColonCA` vs `Fever` (mean r=0.53). These results suggested that learners tended to preserve relative strengths and weaknesses across competencies across cases, but that profile stability varied by case pairing.

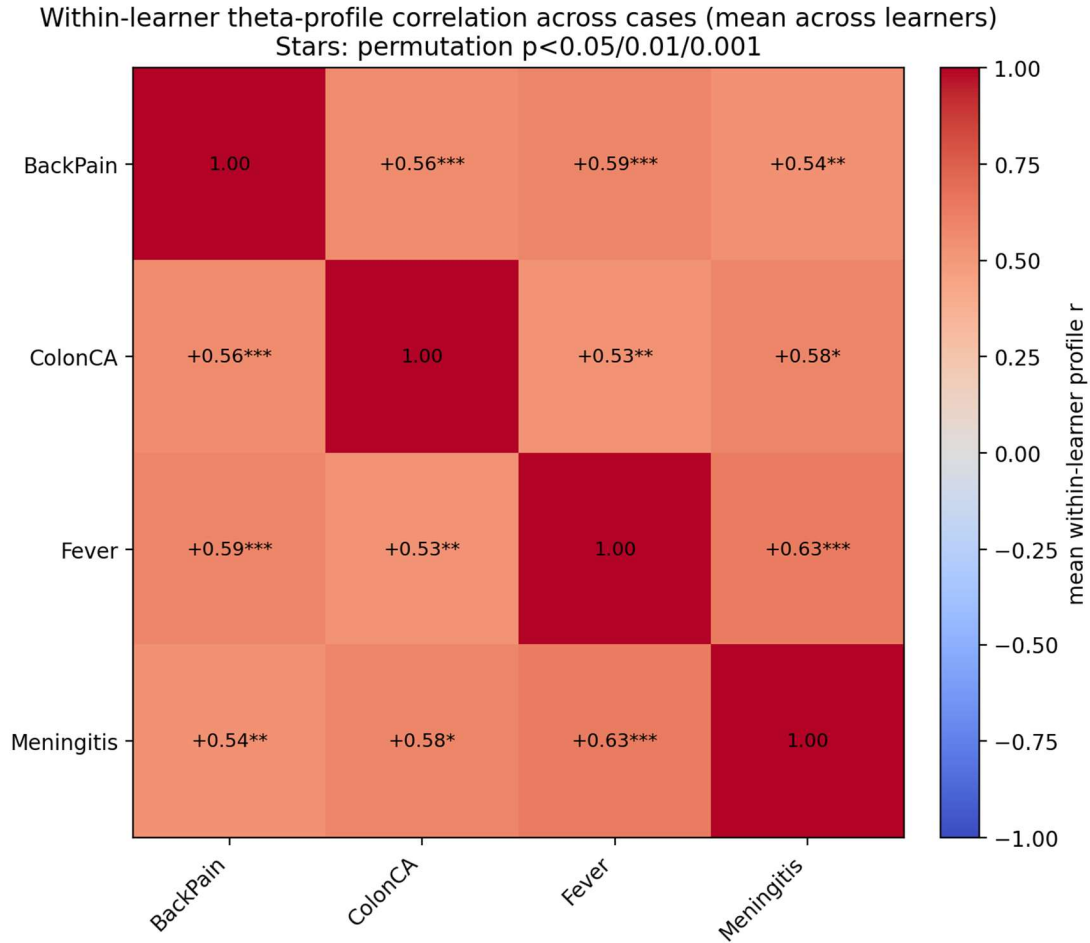

*Figure 10. Within-learner cross-case profile consistency (mean across learners of within-learner profile correlations).*

**Rater effects: theta-dependent detection and criteria.** We summarized theta-dependent rater behavior in the SDT layer. *Detection* shifts were modeled as log-multiplicative effects on sensitivity; *criteria* shifts were additive, with positive values indicating stricter thresholds (sometimes referred to as *severity*,[29] or more conservative use of higher categories) and negative values indicating *leniency*. Across raters, theta-dependent detection shifts (Figure 11) were small in magnitude (approximately within ±0.11 on the log scale), suggesting broadly similar detection across competency dimensions. In contrast, criteria shifts (Figure 12) were larger (approximately spanning about -1.33 to +1.18), indicating that rater severity/leniency varied more substantially by competency dimension; in this run, the largest positive shifts tended to occur for Theta1 (PC) and the largest negative shifts for Theta6 (SBP).

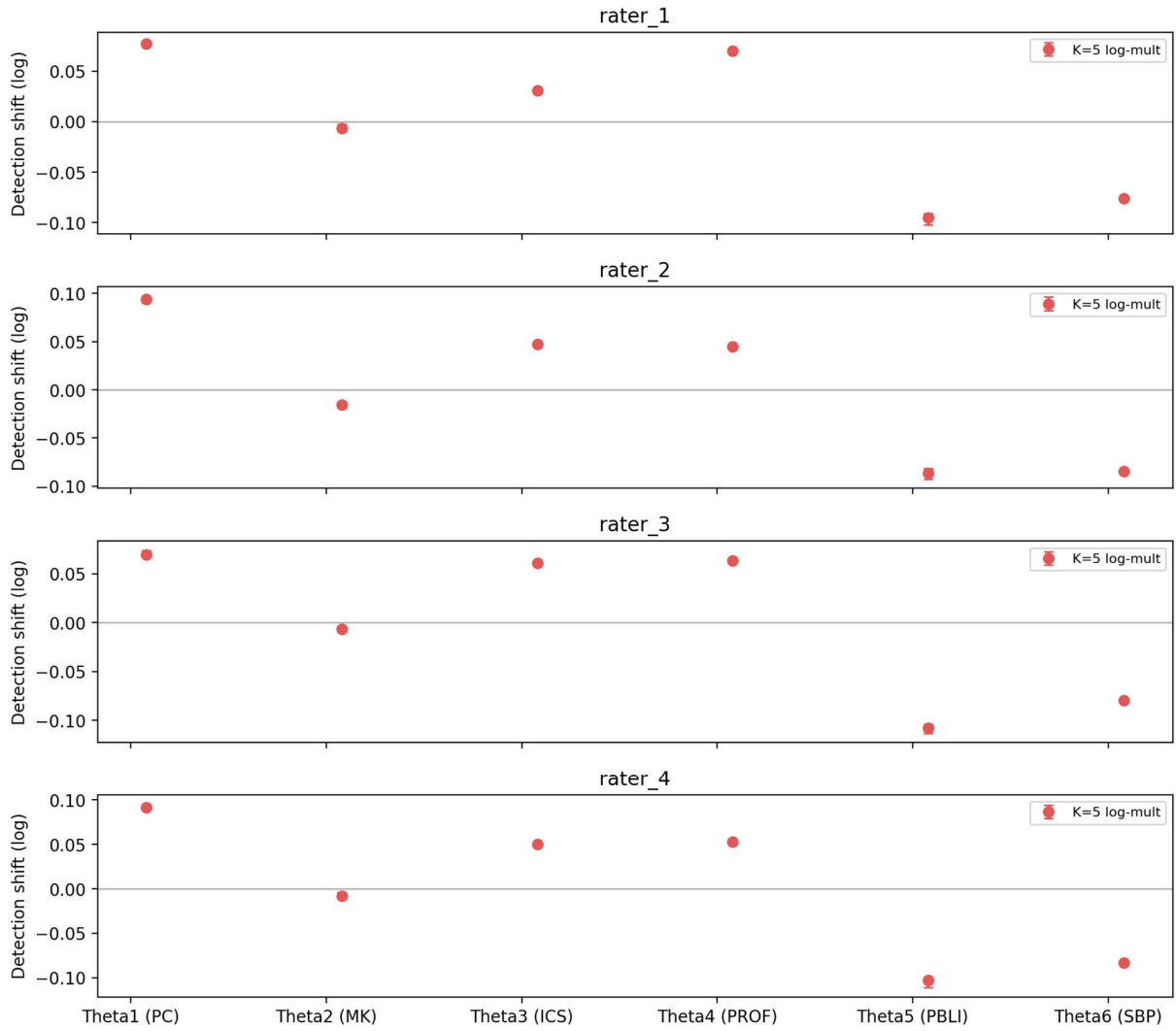

*Figure 11. Theta-dependent detection shifts by rater (log-multiplicative), indicating differences in sensitivity.*

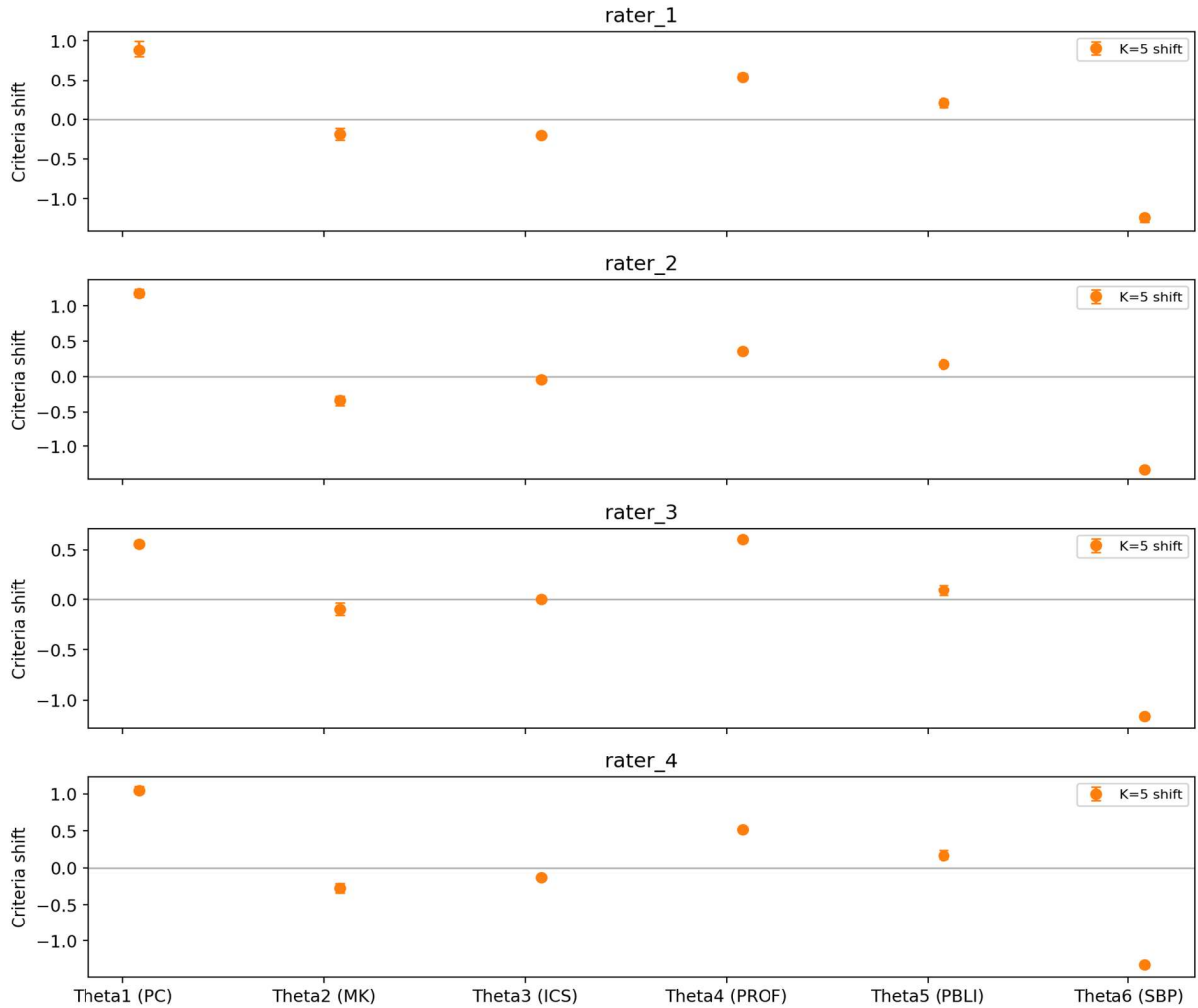

*Figure 12. Theta-dependent criteria (or threshold) shifts by rater (additive), indicating differences in severity/leniency.*

**Item characteristic curve examples.** Item characteristic summaries, based on projecting multidimensional competency onto an item-aligned coordinate, provided qualitative checks on category separation and practical informativeness. An informative item exhibited ordered category transitions with substantial probability mass in interior categories across the ability range (Figure 13). By contrast, near-ceiling and near-floor items (Figures 14 and 15) were dominated by the highest or lowest categories, respectively, and thus provided less information for discriminating learners.

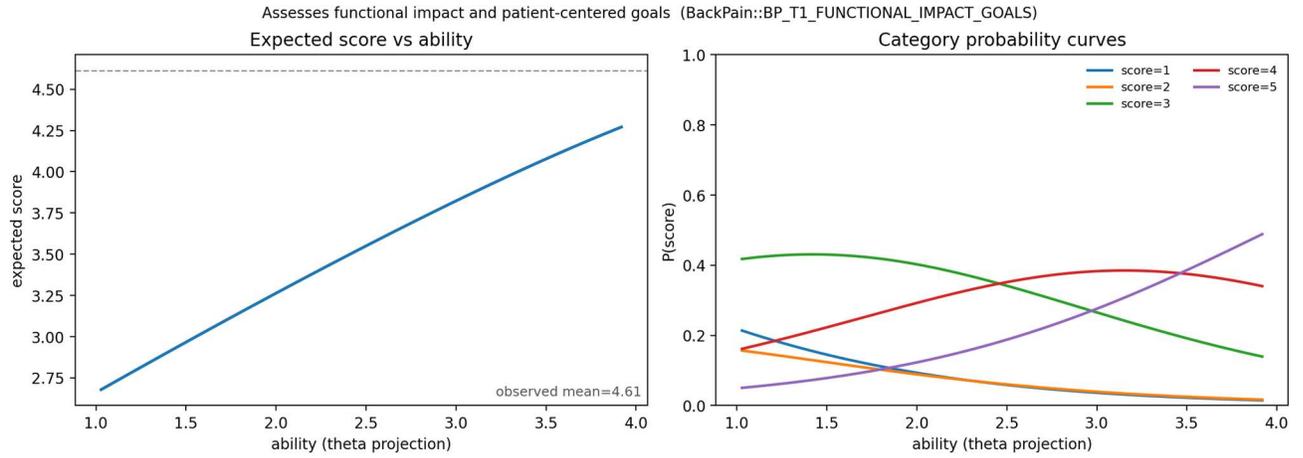

*Figure 13. Example item with informative interior-category use.*

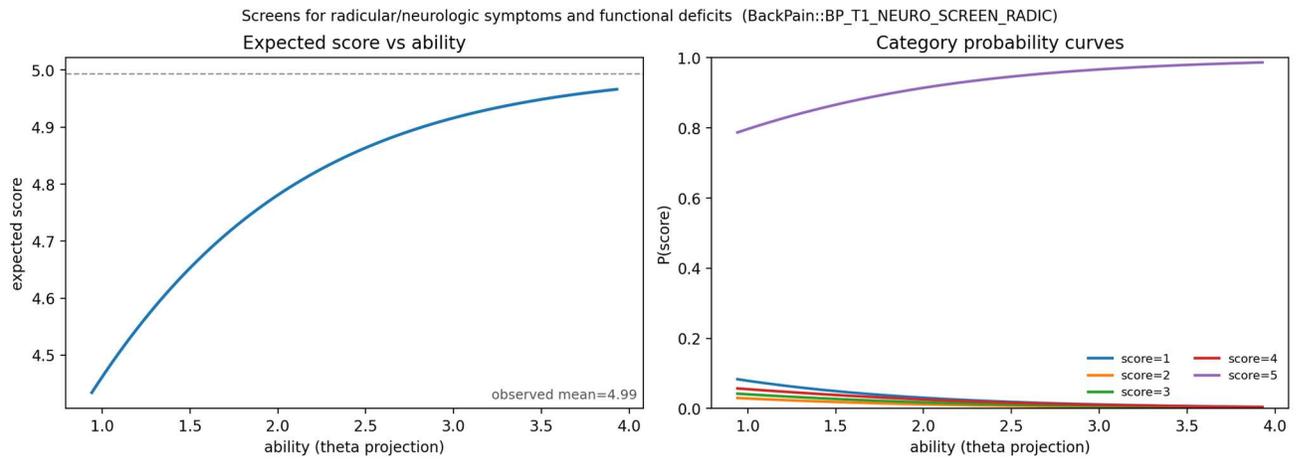

*Figure 14. Example near-ceiling item.*

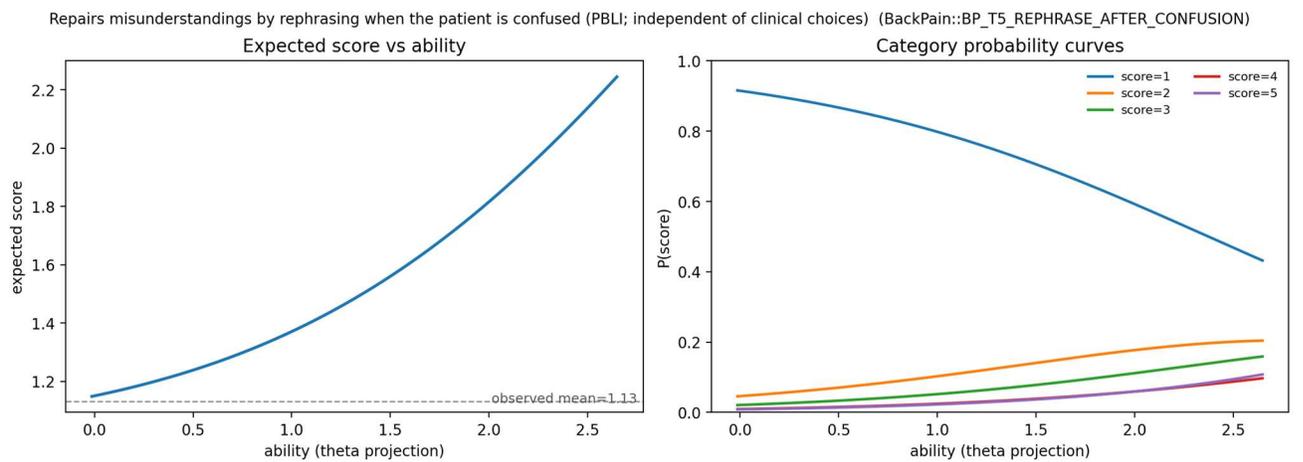

*Figure 15. Example near-floor item.*

**Discussion:**

This study had three aims. First, we aimed to develop and evaluate a measurement framework that estimated learners' clinical competencies on a common scale across virtual patient cases. The results provided substantial support for this aim: competencies were recoverable from end to end in the AI-AI simulation (dimension-wise recovery correlations ranged from r=0.38 to r=0.76), and some competencies showed strong cross-case stability across learners (e.g., Theta4 (PROF) had the highest mean cross-case correlation, r=0.65). However, other competencies exhibited weak cross-case consistency (e.g., Theta2 (MK), r=0.03), suggesting that not all dimensions were equally identified under the current item coverage and case design.

Second, we aimed to estimate case difficulty by competency. The case-by-theta effects (Figures 6-7) demonstrated that cases differed systematically in the degree to which they shifted performance along each competency dimension. Descriptively, `Meningitis` tended to be the most difficult overall (mean case shift -0.96), whereas `ColonCA` tended to be the easiest (mean case shift 0.92). This supported the intended use of the framework for diagnosing which cases elicited and challenged specific competencies.

Third, we aimed to estimate competency-specific rater tendencies (detection and criteria). The theta-dependent SDT summaries (Figures 11-12) demonstrated that rater detection and criteria could be decomposed into baseline rater effects plus competency-specific shifts. In this run, theta-dependent detection shifts were comparatively small, while criteria shifts were larger, indicating that rater criteria (severity/leniency) varied more than rater detection (sensitivity) across competency groups. This supported the feasibility of adjusting for systematic rater differences attributable to rater prompts/models or competency-linked rating behavior.

**Validity evidence**

Interpreted through the AERA 2014 validity framework,[31,32] our results primarily contributed evidence about internal structure and (simulation-based) relations to other variables. Internal structure evidence came from the ability of the HRM-SDT model to recover known competencies, to separate case effects from learner effects, and to characterize rater effects within the SDT layer. Relations to other variables were available because ground-truth competencies were known by design, enabling direct assessment of recovery.

At the same time, content and response-process validity were only partially addressed. Although the simulated cases and rubrics were designed to reflect clinically plausible tasks, we did not yet establish whether the AI learners' interaction strategies matched authentic learner reasoning processes, nor whether the AI rater scoring process matched expert interpretation. Consequential validity (impact on learners, fairness concerns in real use) was intentionally deferred to later phases where humans participated.

A related diagnostic was the correlation between theta dimensions across learners. Because the simulated learner profiles were generated without deliberate cross-dimension correlations, large off-diagonal correlations were not expected in the truth. The estimated theta-correlation matrix exhibited larger induced correlations (max |r|=0.54) than the truth correlation matrix (max |r|=0.34). This suggested that some competencies (ICS and PROF) were not fully separable under

the current measurement design. This finding is not unexpected, since ICS and PROF are closely related constructs that both rely on interpersonal and communications skills.[33–35]

**A blueprint towards learner deployment**

Based on our experience with this AI-AI simulation, we developed a staged, entrustment-based[22] blueprint for deploying learner-facing AI tools in clinical education settings (Figure 16). This blueprint organizes deployment as a progressive sequence of *entrustment phases* – increasing levels of trust in the AI tool that are reflected in increased human involvement and participant risk. Trust is to be based on accumulation of validity evidence specific to each entrustment phase (Figure 17). In Phase 1 (AI-AI), the goal is to establish that the platform produces accurate, relevant, and unbiased content and that the measurement model estimates competence fairly and consistently without putting humans at risk. Subsequent phases incrementally introduce human experts, then advanced learners, then broader learner cohorts, with increasing attention to human fairness, usability, and consequences.

Within this blueprint, the results from Phase 1 serve as a technical readiness screen: they demonstrate that the end-to-end pipeline can recover known competencies, estimate case-by-competency shifts, and estimate rater tendencies, while also revealing specific weaknesses (e.g., low cross-case stability for some competencies) that should be addressed before scaling risk.

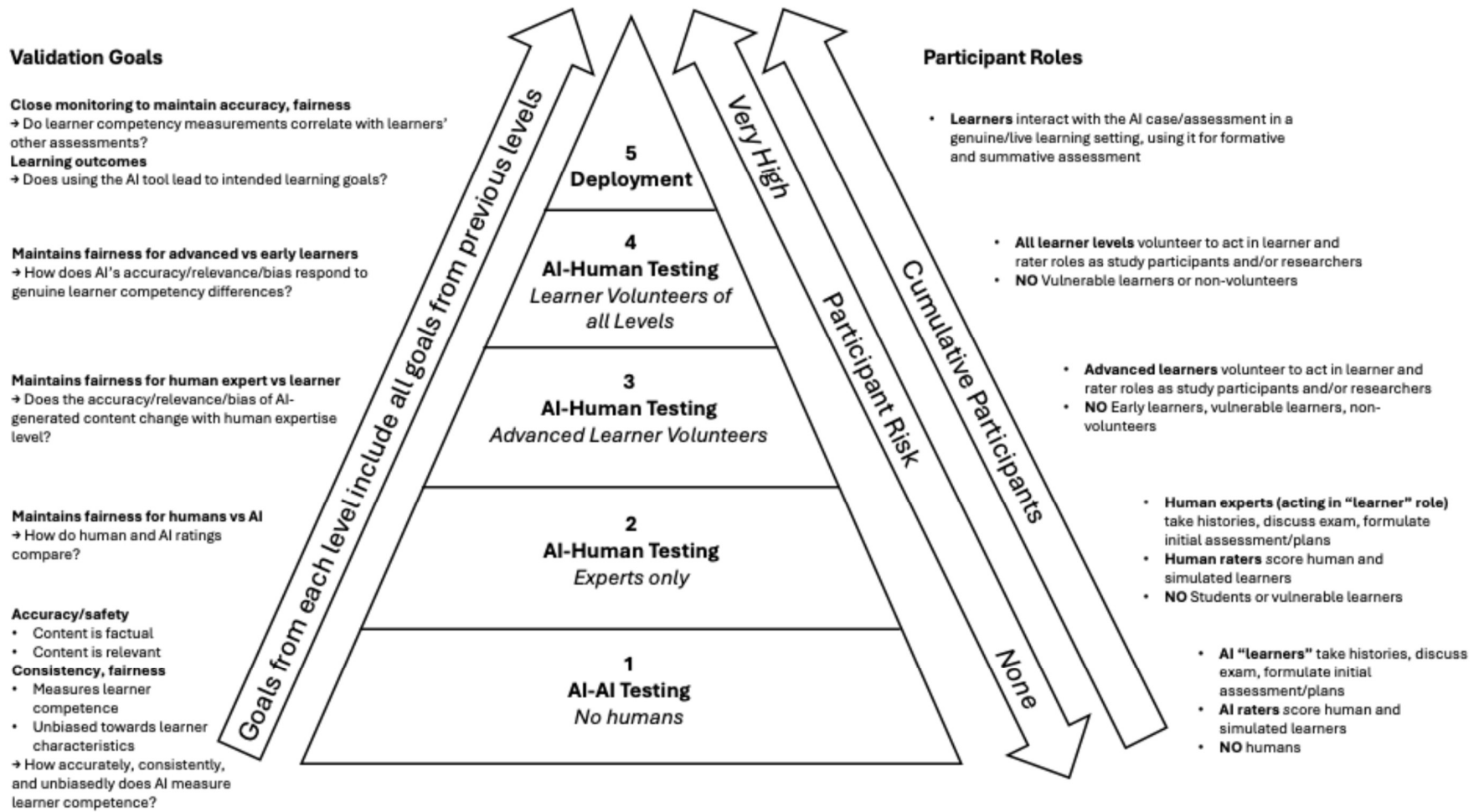

*Figure 16: An entrustment-based blueprint for staged human and learner participation in the co-development of AI-based clinical scenarios and assessment tools*

## Validity Considerations

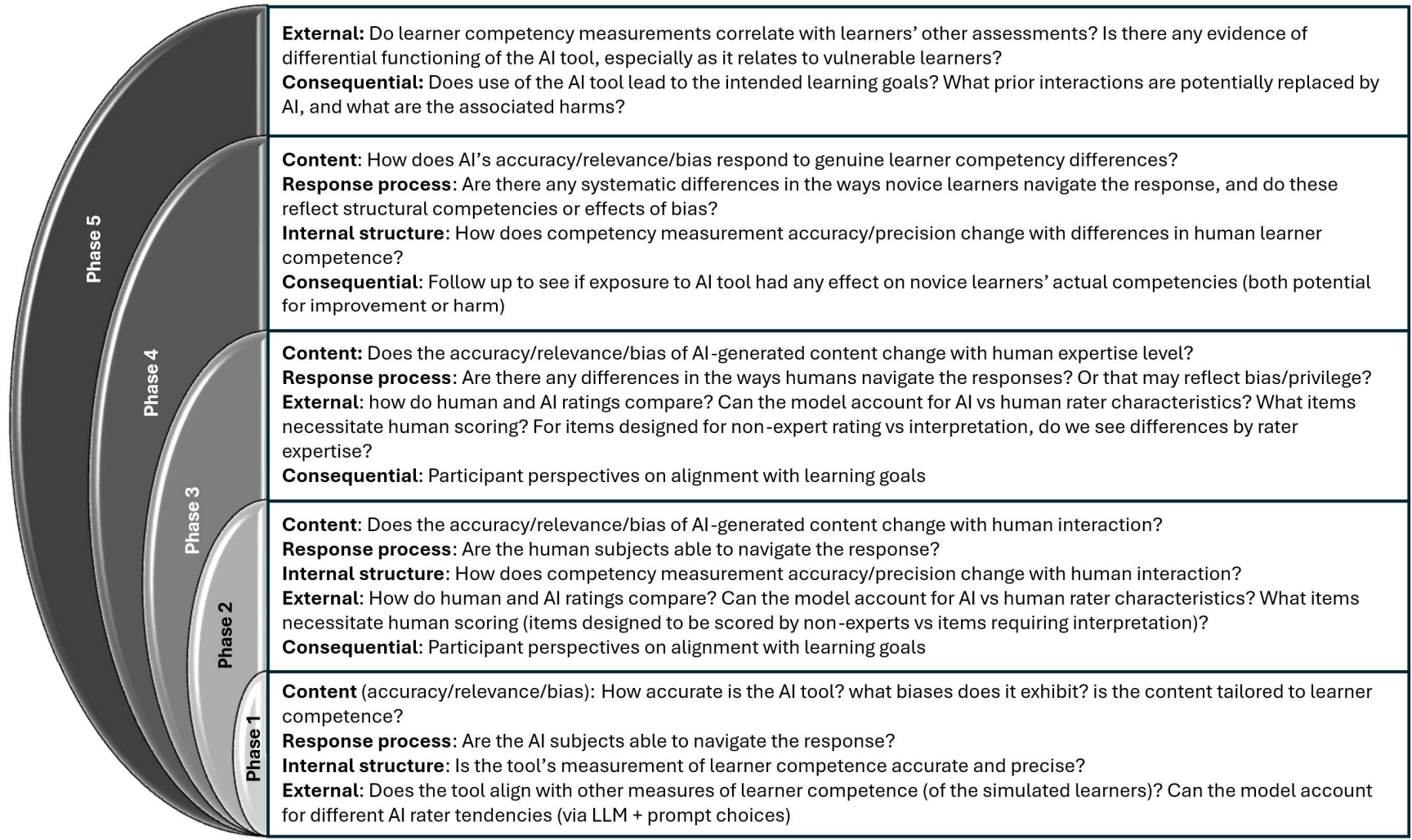

*Figure 17: Validity considerations accumulating in each progressive entrustment phase of the AI deployment blueprint.*

**Validity considerations by entrustment phase**

In Figure 17, we map each entrustment phase to the relevant validity questions that can be answered in that phase. In Phase 1, the emphasis is on content (accuracy/relevance/bias), response process (whether simulated agents can navigate tasks), internal structure (measurement accuracy/precision), and external comparisons (whether competency estimates recover the expected AI learner profiles). Phases 2-5 will expand the scope to include human response processes, agreement between human and AI raters, differential item functioning and fairness across human subgroups, and consequential outcomes in authentic learning environments.

**Limitations and future directions: readiness for phase 2**

Taken together, our Phase 1 results support moving to Phase 2 data collection with human experts, with two important caveats. First, the observed weak cross-case consistency for some competencies indicated that the case set and item blueprint do not yet provide uniformly strong information across all intended dimensions. Second, induced correlations between estimated theta dimensions suggest incomplete separability for some competencies. These caveats do not necessarily preclude Phase 2, but they can inform what Phase 2 should prioritize.

In Phase 2, we will primarily look for (i) content fidelity under human interaction (clinical plausibility, safety, bias), (ii) response-process evidence (whether human experts can navigate the encounters as intended and whether transcripts reflect recognizable clinical reasoning and communication behaviors), (iii) measurement comparability (agreement and systematic differences between expert human ratings and AI rater ratings; calibration of rater detection/criteria across rater types), and (iv) internal-structure robustness (whether cross-case competency stability improves or degrades when the data are generated by humans rather than simulated learners). Human expert data will also allow targeted refinement: identifying which items require expert interpretation, which can be rated reliably by non-experts, and which competency dimensions require additional or redesigned items to achieve stronger separability and case invariance.

## Conclusions:

Our overall vision is to crowdsource the development of an open-source virtual patient platform that can reliably generalize across worldwide contexts – allowing learners to interact and receive feedback from with virtual patient cases representing a global case-mix. While the barrier to authoring clinically-sound cases for use with LLM-enacted standardized patients is relatively low, validating cases is a more complicated task. While laborious, a minimum validation requirement should stand as a safeguard before allowing learners to interact with each case. Here we have developed a measurement model and validation blueprint that can help guide this process and minimize risk to learners (Figures 16 and 17). We propose validation of cases via this phased process, with each advancement gated by accumulation of key validity evidence at each stage. To provide a quantitative backbone to support these validation steps, our measurement model ties together the critical assessment observables: learner competencies with case, rater, and item variation. In this study, we used this model to assess the psychometric properties of virtual patient cases and feedback system prior to any contact with human subjects.

Our validation blueprint and measurement model need not be restricted only to virtual patient-based assessments (with AI in patient and rater/feedback roles), but they can also be applied to learning scenarios where humans assume those roles as well, including: standardized patient-based assessments, OSCE scoring/feedback, ambient feedback on actual patient encounters, and real time AI nudges systems. Additionally, once the model parameters for cases, raters, and items have been estimated (with adequate certainty for a relevant learner population), they can be generalized for rapid estimation of learner competencies without needing to re-fit the entire model. We have used such pre-estimated parameters to implement real-time scoring of learner competencies during cases.[b]

We plan to proceed with our proposed validation blueprint via a participant-as-researcher, crowd-sourced approach. To maximize transparency in this process of co-development, we plan to publish our results from each phase of our validation, and to provide open access to our platform so that the HPE community may also join is in this crowdsourced approach.

---

[b] The model can provide estimates mid-case even without all items being scored, afforded by the item opportunity gating mechanism.